\documentclass[12pt]{article}
\usepackage{pdproc}
\usepackage{units} 
\usepackage[pdftex]{graphicx}

\newcommand{\pot}{\ensuremath{3.14\times10^{20} {\rm \,POT}}}
\newcommand{\mcnnpred}{\ensuremath{22\pm 5 {\rm (stat.)} \pm 3 {\rm (syst.)}}}
\newcommand{\annpred}{\ensuremath{27\pm 5 {\rm (stat.)} \pm 2 {\rm (syst.)}}}
\newcommand{\mcnnobs}{\ensuremath{28}}
\newcommand{\annobs}{\ensuremath{35}}

\newcommand{\delmsq}[1]{\ensuremath{\Delta m^2_{ #1 }}}
\newcommand{\dmsqx}[1]{\delmsq{ #1 }}
\newcommand{\sinsq}[1]{\ensuremath{\sin^{2}\left(2\theta_{ #1 }\right)}}

\newcommand{\numu}{$\nu_{\mu}$}

\newcommand{\nue}{$\nu_{e}$}

\newcommand{\nutau}{$\nu_{\tau}$}
\newcommand{\dmsqtwo}{$|\Delta m^{2}_{32}|$}

\newcommand{\dmsqonetwo}{$|\Delta m^{2}_{21}|$}

\begin{document}

\renewcommand{\thefootnote}{\alph{footnote}}
  
\title{Recent Results from the MINOS experiment 
}

\author{Milind V. Diwan}

\address{ 
Physics Department, Brookhaven National Laboratory \\
Upton, NY 11973, USA \\
 {\rm E-mail: diwan@bnl.gov}\\ 
April 20, 2009 
}

\abstract{MINOS is an accelerator neutrino oscillation experiment 
at Fermilab. An intense high energy neutrino beam is produced at Fermilab
and sent to a near detector on the Fermilab site  and also to a 5 kTon
 far detector 735 km away in the Soudan mine in northern
 Minnesota. The experiment has now had several years of running with 
millions of events in the near detector and 
hundreds of events recorded in the far detector. 
I will report on the recent results from this experiment
which include precise measurement of \dmsqtwo, analysis of 
neutral current data to limit the component of sterile neutrinos, and 
the search for $\nu_\mu \to \nu_e$ conversion.  
The  focus will be on the analysis of data for 
$\nu_\mu \to \nu_e$ conversion.
Using data from an exposure of $3.14\times 10^{20}$ protons on target, we
have selected electron type events in both the near and the far detector.
The near detector is used to measure the background which is extrapolated to 
the far detector. We have found 35 events in the signal region with a background 
expectation of $27\pm 5(stat)\pm 2(syst)$. Using this observation we set a 
$90\%$ C.L. limit of $\sin^2 2 \theta_{13} < 0.29$ for 
$\delta_{cp} = 0$ and 
normal mass hierarchy.  
Further analysis is under way to reduce backgrounds and 
improve sensitivity. }   
\normalsize\baselineskip=15pt

\section{Introduction}
In the current picture of neutrino oscillations, three flavors of 
neutrinos are related to three mass states by 
the Pontecorvo-Maki-Nakagawa-Sakata mixing matrix
\cite{ref:pmns,maki}. 
The mixing can be described by two $\Delta m^2$ parameters (\dmsqtwo, 
\dmsqonetwo),
 three mixing angles ($\theta_{23}$, $\theta_{12}$, and $\theta_{13}$) and a
 CP violating phase ($\delta_{cp}$)\cite{ref:pdg}.
The oscillation phenomena naturally falls into two domains: 
the  atmospheric neutrino oscillations,   
and Solar neutrino oscillations \cite{ref:pdg}. The atmospheric neutrino 
oscillations  are  
  well-described by $\nu_\mu \rightarrow \nu_\tau$ oscillations,
with parameters $\sin^2 2\theta_{23} > 0.92$ 
and $1.9\times 10^{-3} < |\Delta m_{32}^2| < 3.0\times 10^{-3}~{\rm eV}^2$
at 90$\%$ C.L \cite{ref:osc1}. The K2K experiment and MINOS have confirmed
 the atmospheric 
neutrino oscillations with accelerator beams 
\cite{ref:osc5,ref:minosprl} 
Solar $\nu_e \rightarrow \nu_{\mu,\tau}$ oscillations 
are described by  $\sin^2 2\theta_{12}= 0.86^{+0.03}_{-0.04}$
 and $ \Delta m_{21}^2 = 8.0^{+0.4}_{-0.3}\times 10^{-5}~{\rm eV}^2$ 
are also consistent with multiple observations \cite{ref:osc6,ref:osc7}, 
and are confirmed by disappearance of reactor 
$\bar{\nu}_e$\cite{ref:osc8}.
As yet, very little is known about either 
$\theta_{13}$ or $\delta_{cp}$, although lack of observed 
disappearance of
 reactor $\bar{\nu}_e$ over a few km baseline\cite{Apollonio:2002gd}
 has shown that $\theta_{13}$ must be small: $\sin^2 2 \theta_{13}<0.19$ at 
$90\%$ C.L. 
Furthermore, the sign of \dmsqtwo ~(or the ordering of the mass eigenstates) 
is unknown. The sign of $\Delta m^2_{12}$ ~is known using the strong 
matter effects that must be considered when analyzing neutrinos from 
the Sun.  Determination of the unknowns in neutrino mixing needs
further experiments in the oscillation range of 
\dmsqtwo ~for conversion of muon and electron type neutrinos into each 
other.

For our present
discussion, it is useful to exhibit an approximate analytic formula for the
oscillation of $\nu_\mu \to \nu_e$ for 3-generation mixing obtained with the
simplifying assumption of constant matter density \cite{freund,cervera}. 
Assuming a constant matter density, 
the oscillation of $\nu_{\mu} \rightarrow \nu_e$ in the Earth 
for 3-generation mixing is described
approximately by  Equation \ref{qe1}.
In this equation $\alpha=\Delta m^2_{21}/\Delta m^2_{31}$, $\Delta = \Delta
m^2_{31} L/4E$, $\hat{A}=2 V E/\Delta m^2_{31}$,
$V=\sqrt{2} G_F n_e$. $n_e$ is the density of electrons in the Earth. 
Recall that $\Delta m^2_{31} = \Delta m^2_{32}+\Delta m^2_{21}$. 
Also notice that $\hat{A}\Delta$, which has absolute value of $ L G_{F} n_e/\sqrt{2}$, is sensitive  
to the sign of $\Delta m^2_{31}$. 

%\begin{widetext}
\begin{eqnarray}
P(\nu_{\mu} \rightarrow \nu_e) &\approx&
\sin^2 \theta_{23} {\sin^2 2 \theta_{13}\over (\hat{A}-1)^2}\sin^2((\hat{A}-1)\Delta)  \nonumber\\ &&
+\alpha{\sin\delta_{CP}\cos\theta_{13}\sin 2 \theta_{12} \sin 2
\theta_{13}\sin 2 \theta_{23}\over
\hat{A}(1-\hat{A})} \sin(\Delta)\sin(\hat{A}\Delta)\sin((1-\hat{A})\Delta) \nonumber\\ &&
+\alpha{\cos\delta_{CP}\cos\theta_{13}\sin 2 \theta_{12} \sin 2
\theta_{13}\sin 2 \theta_{23}\over
\hat{A}(1-\hat{A})} \cos(\Delta)\sin(\hat{A}\Delta)\sin((1-\hat{A})\Delta) \nonumber\\ &&
+\alpha^2 {\cos^2\theta_{23}\sin^2 2 \theta_{12}\over
\hat{A}^2}\sin^2(\hat{A}\Delta) \nonumber \\
\label{qe1}
\end{eqnarray}
%\end{widetext}

For anti-neutrinos, the second term in Equation \ref{qe1}  
has the opposite sign. It proportional to the CP violating
quantity $\sin \delta_{CP}$. 
An accelerator experiment using high energy neutrinos, a sufficiently 
long baseline, and the ability to detect $\nu_\mu \to \nu_e$ conversion 
with low backgrounds and high statistics,  
has sensitivity to all four terms in Equation \ref{qe1}. 
The first term dominates the sensitivity to the unknown parameter 
$\theta_{13}$. MINOS is the first high energy 
accelerator experiment so far to have sensitivity to more than 
the first  term in Equation \ref{qe1}. 
In the following we describe an analysis of the MINOS data to 
reduce backgrounds to allow the observation of  
$\nu_\mu \to \nu_e$.

\section{MINOS beam and detector}

\begin{figure}
\centering\leavevmode
\includegraphics[angle=0,width=0.7\textwidth]{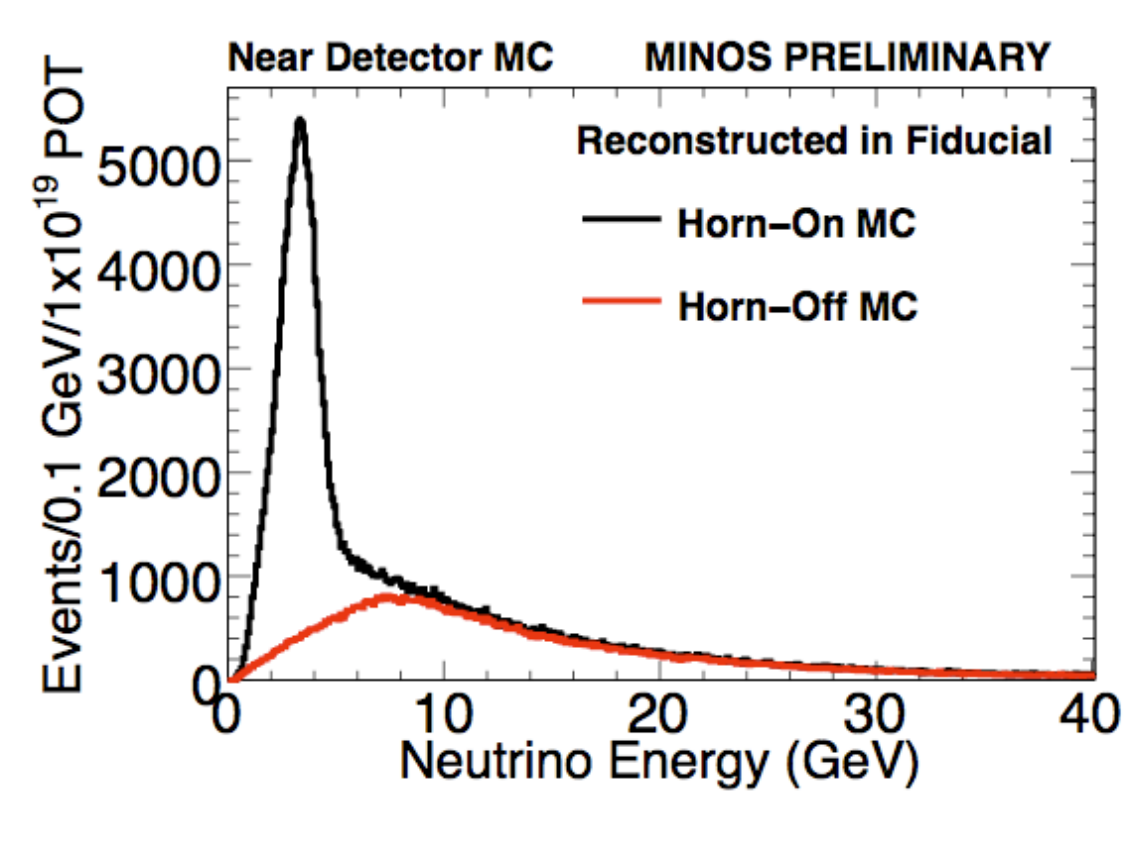}
 \caption{(in color)
Spectrum of all muon neutrino  events in the MINOS near detector fiducial volume
for the horn on and horn off configurations.   
  \label{nearspec} }
\end{figure}

The MINOS detectors \cite{ref:minos} and the NuMI beam line \cite{ref:numi}
 are described elsewhere. 
 In brief, 
NuMI is a conventional two-horn-focused neutrino beam with a 
675~m long decay tunnel.  The neutrino 
 beam goes through the Earth to upper Minnesota 
over a distance of 735 km to the far detector.  
The horn current and position of the hadron production target relative to the
 horns can be configured to produce different $\nu_\mu$ energy spectra.  
In figure \ref{nearspec} we show the spectrum of all $\nu_\mu$ 
events in the fiducial volume for the horn-on and horn-off configurations. 
 Several beam configurations with different mean energies have been 
used for studying backgrounds and systematics: in particular, the horn-off 
configuration has
 been particularly useful for the $\nu_e$ search. The high energy, 
$\sim 5-10 GeV$, obtained by moving the production target,
 as well as intermediate energy configurations 
 have been used for the analysis of the beam systematics for the 
muon neutrino disappearance data. 
Most of the physics data has been in the low energy (horn-on) 
configuration in which the 
peak of the spectrum is $\sim 3 GeV$. In the low energy configuration, $92.9\%$
 of the 
flux is $\nu_\mu$, $5.8\%$ is $\bar\nu_\mu$ and $1.3\%$ is the $\nu_e/\bar\nu_e$ 
contamination.

MINOS consists of two detectors: a 0.98~kt
 Near Detector (ND) 1.04~km from the NuMI target; and 
a 5.4~kt Far Detector (FD) 735\,km from the target.  
 Both are segmented, magnetized calorimeters that permit
 particle tracking, optimized for neutrino energy range of 
 $1 < E_{\nu}<  50 {\rm GeV}$. 
  The curvature of muons produced in
 $\nu_\mu +\mbox{Fe}\rightarrow\mu^-+X$ interactions 
\footnote{Approximately 5\% of the neutrino  interactions 
occur in aluminum and scintillator.} is used 
for energy determination of muons that exit the detector and 
to distinguish the $\nu_\mu$ component of the
 beam from the contamination.  
  The energy of muons contained in the detector
 is measured by their range.
The muon and shower energies are added 
to obtain the 
reconstructed muon neutrino energy ($E_{reco} = E_\mu + E_h$)  with a 
resolution given by  
$\Delta p_\mu/p_\mu \approx 10\%$ and 
$\Delta E_h/E_h \approx 56\%/\sqrt{E_h}$. 
  For electron neutrino 
detection the relevant parameters of the detector 
concern the calorimetric segmentation. Both the near and far detectors 
have identical segmentation of 1 inch (1.44 radiation length) steel 
and 1 cm thick plastic scintillator.  Transversely the scintillator 
is in strips of 4.1 cm width, corresponding to Moliere radius of 3.7 for 
electromagnetic showers.  The scintillator is read by wavelength 
shifting fibers into multianode PMTs.  
The scintillator strips range in 
length from a maximum of 8 meters in the far detector 
 down to $\sim$ 1 m in the near detector.  
The light yield is on the average $\sim$6 
photo-electrons for a minimum ionizing 
particles. Although the near and far detectors have identical granularity,
there are important differences: the light yield, the type 
of  PMT used (Hamamtsu M16  in
the far, and Hamamatsu M64 in the near), the cross talk between channels in 
the PMTs, and multiplexing of scintillator strips onto the PMT pixels
\cite{ref:minos}. 
These differences are carefully calibrated using cosmic rays
and  simulated in the Monte Carlo programs 
to limit the near and far differences. After electron particle identification 
and selection cuts as described below, the energy resolution for 
$\nu_e$ events is $\sim 30\%/\sqrt{E}$.

\section{Data Reduction}

This note describes results from data 
recorded between May 2005, and July 2007.
 Over this period, a total of $3.36\times10^{20}$~protons on target
 (POT) were accumulated.
  A $1.27\times10^{20}$~
POT subset of this exposure (hereafter referred to as Run~I) 
forms the data set  from Ref~\cite{ref:minosprl}. 
 In Run~I and for most of the new running period (Run II),
 the  beam line was configured to enhance $\nu_\mu$ 
production with energies 1-5~GeV (the low-energy configuration).
  An exposure of $0.15\times10^{20}$~POT was accumulated 
with the beam line configured to enhance the
 $\nu_\mu$ energy spectrum at 5-10\,GeV (the 
high-energy configuration).  The Run II data were
 collected with a replacement target of identical
 construction due to failure of the motion system of the first target. 
 The new target was found to be displaced 
longitudinally $\sim$1~cm relative to the first target, resulting in a
 30 MeV shift in the neutrino spectrum.  
This effect is incorporated in the Monte Carlo simulation, and the
 Run I and Run II data sets are analyzed separately to account for this shift.

The data reduction has several components: cuts are first applied to 
remove data from periods of bad detector and beam conditions. After event 
reconstruction, 
preselection cuts  select events that are enriched in the types 
of events that are under analysis: negative or positive muons, 
electrons or neutral currents.
These cuts are performed as identically as possible for the near and far 
detectors. Differences are accounted for in the Monte Carlo. 
After the preselection, particle identification cuts are applied 
to extract a pure sample  of the events under consideration. 
The muon and neutral current analysis has been described in 
detail in previous publications \cite{ref:minosprd,ref:minosnc}. 

For the electron analysis the selected   sample of data 
(the low energy horn on configuration) corresponds to 
$3.14\times 10^{20}$ protons on target for the far detector. 
The near detector data was sampled uniformly and scaled to 
correspond to $10^{19}$ protons on target.  
Reconstructed events were chosen within the well calibrated 
parts of the detectors corresponding to fiducial masses of 
29 ton and 4 kton for the near and far detectors, respectively,
within the 10 $\mu$sec beam pulse gate to reject  
cosmic ray events. After these cuts cosmics  contribute $<0.5$ event background in the final 
sample.    
The $\nu_e$ preselection cuts selected an initial 
sample of events with single 
electromagnetic showers according to the 
reconstruction algorithm and rejected 
events with any tracks longer than 25 planes. A second cut examined 
planes with track-like hits, and eliminated events with more than 
 16 planes with such hits.   
After the precuts, the event sample is composed of 
$\sim 30\%$ $\nu_\mu$ CC events in which the muon is too short to be 
rejected, $\sim 65\%$ NC events,
and about $\sim 5\%$ $\nu_e$ events from the beam
contamination.  

\section{Selection of Electron Neutrinos}

After preselection, further rejection of neutral current, and muon charged 
current events is needed. To achieve this rejection we use the short 
compact nature of the electromagnetic showers compared to the diffuse nature of 
hadronic showers. We have developed two software algorithms to examine 
each candidate event and classify it as a potential electron neutrino signal 
or background. 

\underline{Selection ANN} 
We use the pattern of energy deposition, after eliminating hits with less than 
2 photo-electrons,  to characterize each event
by several parameters.  
The artificial neural network (ANN) algorithm combines 11 such reconstructed 
quantities that exhibit signal and background separation.  Some of these 
 quantities are the maximum energy fraction in 4 planes,
fraction of energy in a 3 strip wide road, the RMS of transverse 
energy deposition, etc. 
The output from the ANN is  between 0 
(background like) and 1 (signal-like). With a cut at 0.7, the efficiency for 
the signal, after preselection, 
 is expected to be approximately 41\%; the expected 
neutral current  and  
$\nu_\mu$CC rejection efficiency is
 $\sim 92.3\%$ and 
$\ge 99.4\%$, respectively.

\underline{Selection LEM} The second discrimination technique is a novel approach 
called Library Event Matching (LEM) selection in 
which each event candidate is compared to a large library
of simulated $\nu_e$-CC and NC events. The best 50 library 
matches are found for each candidate event, by considering the probability 
that two different energy deposition 
patterns in the detector originated from the same 
neutrino interaction.
This computationally intensive 
technique can be carried out because the size of the events is
generally small, and all the strip information can be used. 
 Three variables are constructed: the 
fraction of these matches that are $\nu_e$-CC events, the mean 
hadronic $y$ of the best matches, and the mean fractional 
charge $q$ matched within those best matches. A likelihood is then 
formed from these variables as a function of energy.
With a cut at 0.65, the efficiency for 
the signal, after preselection, 
 is expected to be approximately 46\%; the expected 
neutral current  and  
$\nu_\mu$CC rejection efficiency is
 $\sim 92.9\%$ and 
$\ge 99.3\%$, respectively.

 The two selection algorithms rely on very different techniques, 
provide different 
signal to background ratios, and are sensitive to different systematic 
uncertainties. Assuming the signal is  at the Chooz limit, 
LEM has the potential to  achieve a better signal to background ratio
(1:3) compared to ANN (1:4), but it is more sensitive to systematic uncertainties on the 
relative energy calibrations in the near and far detectors.  
Both algorithms select predominantly NC events and higher y, $\nu_\mu$ CC 
events. The background 
consists mainly of deep inelastic scattering events, with
nearly half of the background showers containing a single $\pi^0$. 
In the analysis reported here, the ANN selected sample 
is used to derive the final results, but the LEM 
selection is examined as a cross check.

\section{Calculation of Backgrounds}

\begin{figure}
\centering\leavevmode
\mbox{\includegraphics[angle=0,width=0.49\textwidth]{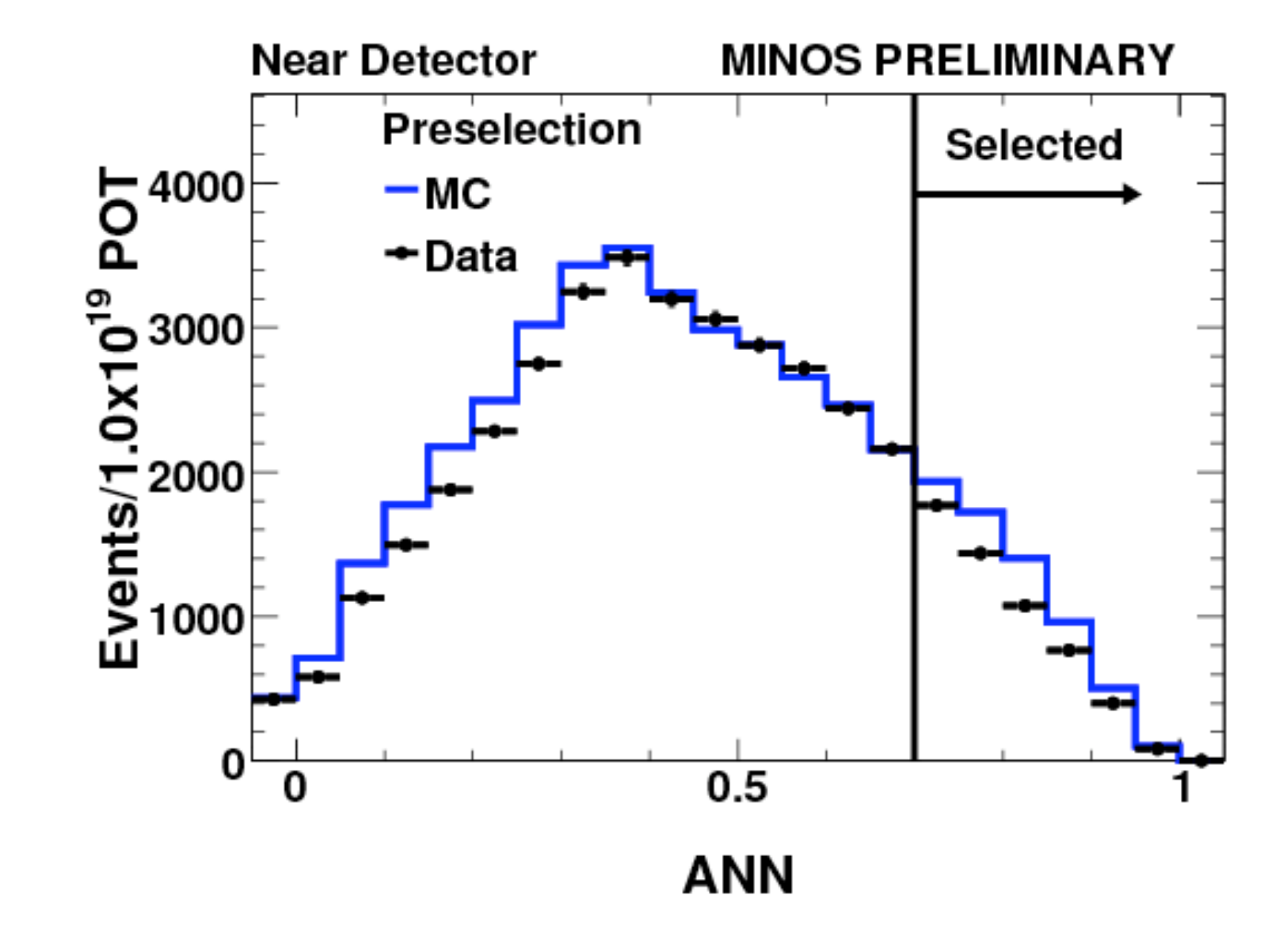}
\includegraphics[angle=0,width=0.49\textwidth]{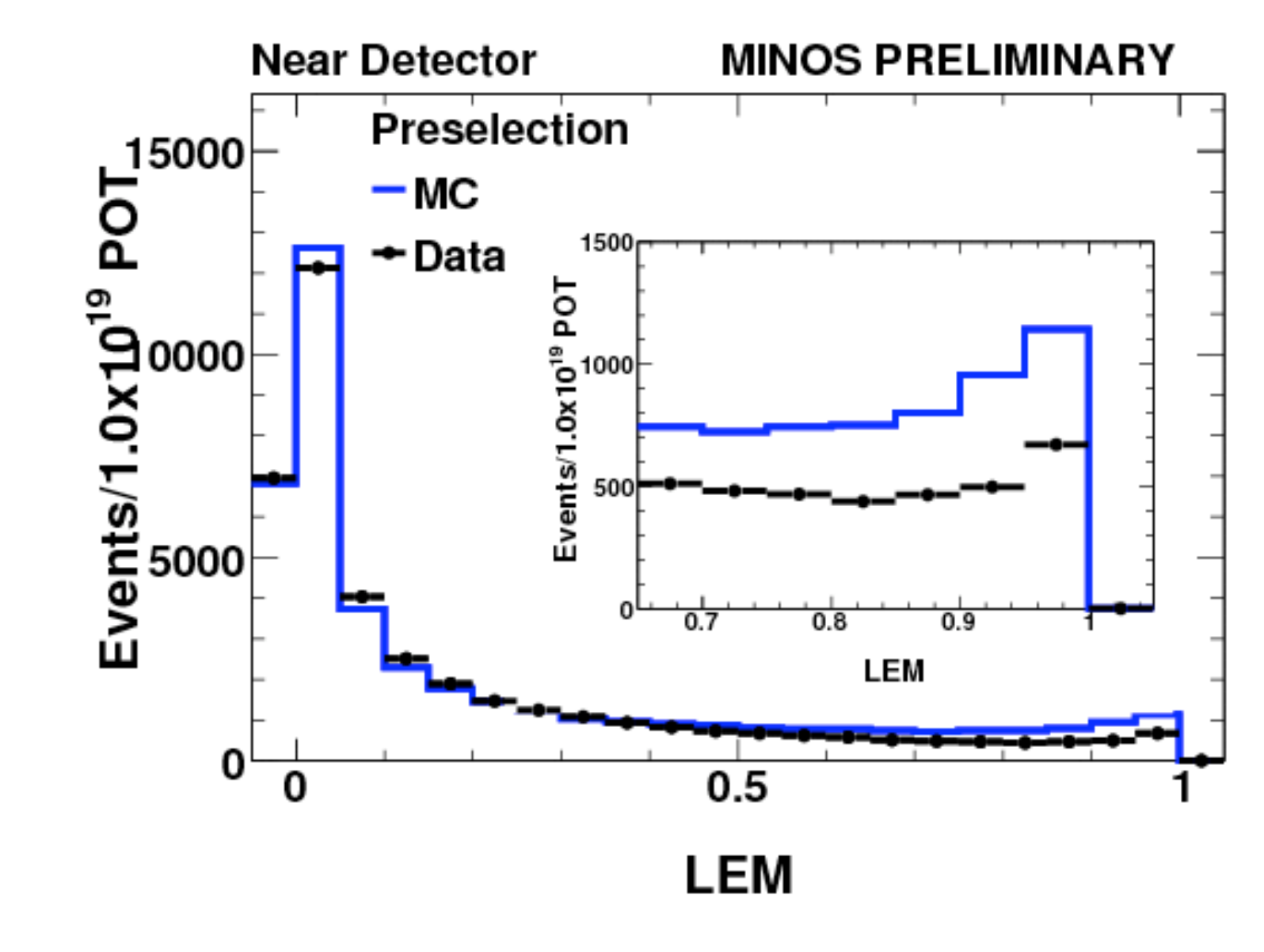}
}
 \caption{(in color)
The ANN PID distribution for the near detector data (left). 
The LEM PID distribution for the near detector data (right). The plots show the 
chosen cuts for selection of $\nu_e$-like events. 
%On the right, 
%comparison of reconstructed energy spectrum of e-CC 
%selected backgrounds in the ND data and MC simulation. 
%Black points show data points with statistical errors. Red histogram shows the MC simulation, 
%with shaded regions indicating systematic uncertainties stemming from hadronic shower modeling.
% Blue histogram shows the NC component of the background as derived from the horn-off method, 
%while (the other) red histogram shows the CC component from the horn-off method.
% Shaded purple histogram shows the beam $\nu_e$-CC component.
  \label{nearpid} }
\end{figure}

 The rate and spectrum of events selected as electron like in the 
near detector are used to predict the number of background 
events expected in the far detector.
Figure \ref{nearpid} shows the distribution 
of ANN PID and LEM PID for the  
near detector data. The plots also show the prediction
 from the Monte Carlo which deviates from the
data. This level of deviation is within the systematic errors due to 
cross section and hadronic shower modeling uncertainties. The
Monte Carlo is based on past data 
\cite{ref:nugen}
with much lower statistics in the MINOS energy region, 
and therefore while the Monte Carlo can be used 
for understanding ratios,  and  relative changes, the MINOS 
near data itself must be used 
to determine the normalization of the background in the far detector.

%The plot on the right hand side shows the energy distribution of 
%the events after the PID cut. 
%The figure shows that the central values of the data and Monte Carlo disagree at 
%the level of 20\%, however this is 
% within the large error bars of the Monte Carlo. These errors result from our estimate 
%of the uncertainties on NC cross sections and hadronic shower modeling. In addition, 
%the particle identification requirement forces hadronic showers with larger electromagnetic 
%fluctuations, which are difficult to model into the selected sample. Examination of the 
%LEM selected events shows even greater disagreement with the simulation at the 40% level.

 The near detector background spectrum has three different components,
 NC events, $\nu_\mu$-CC events and beam $\nu_e$
 events. 
At lowest order the far detector background calculation is 
the near detector event rate (5524 events per $10^{19}$ POT for ANN) 
 multiplied by the   energy averaged far/near ratio of 
the neutrino flux  $\sim 1.3\times 10^{-6}$ and the 
ratio of the fiducial masses $4 {\rm kton}/ 29 {\rm ton}$.  
However, the $\nu_\mu$ charged current component of the background is affected 
by oscillations, and therefore a more sophisticated calculation is needed. 
For such a  calculation we need to 
separate the background components and extrapolate them separately to the 
far detector. 
The far detector will also have a small component from $\nu_\tau$ events
which will be calculated by Monte Carlo.

\begin{figure}
\centering\leavevmode
\mbox{\includegraphics[angle=0,width=0.49\textwidth]{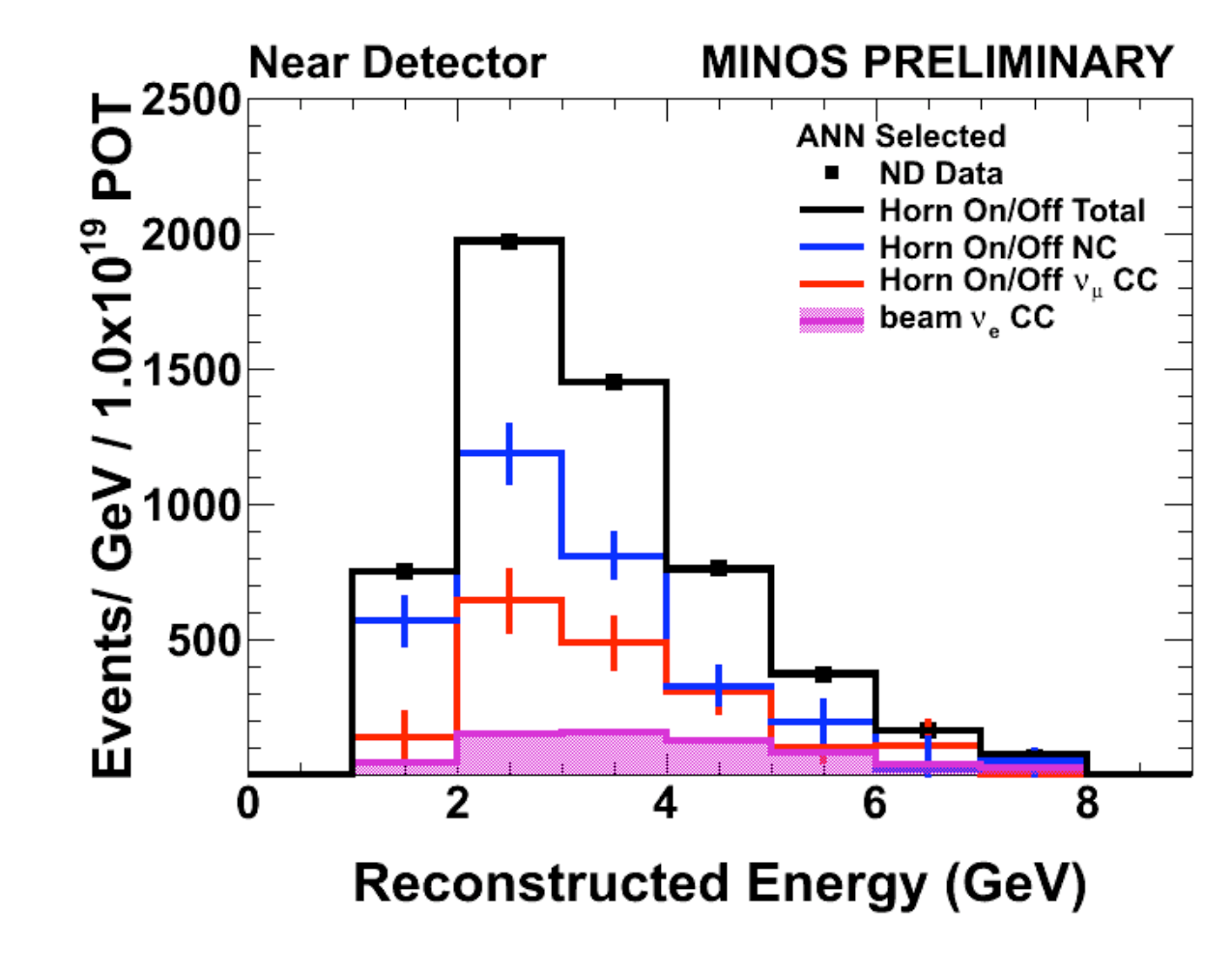}
\includegraphics[angle=0,width=0.49\textwidth]{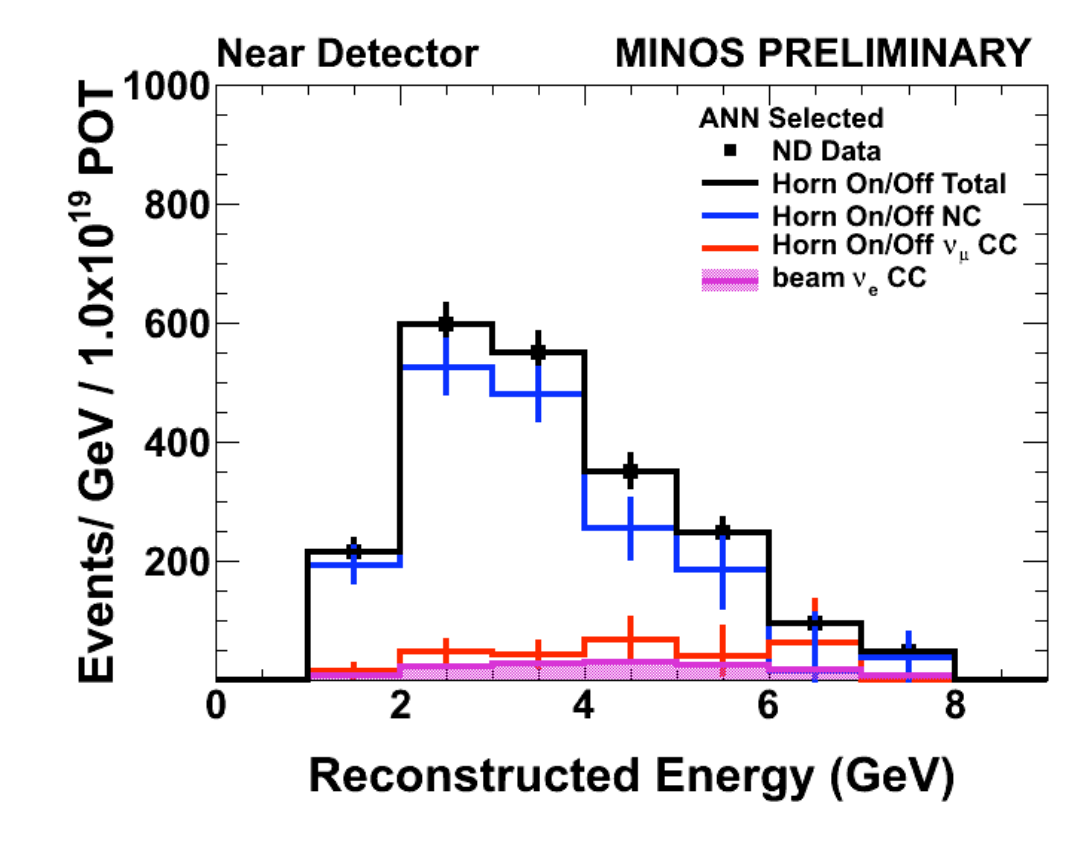}}
 \caption{(in color)
Separation of the types of backgrounds in the near detector data. 
A calculation is performed using the horn-off data which is enriched 
in NC events because of the higher energy neutrino spectrum (see text).
Large fraction of the error is due to the statistics of the horn-off sample.    
This is for the ANN selection, results 
are similar for the alternate selection (LEM).  
  \label{honoff} }
\end{figure}

The components of the background are  determined using the horn off data sample recorded in the ND. 
 Applying the \nue{} selection to data taken with the focusing horns turned off (horn-off)
 provides a neutral current enriched sample.  The higher mean energy spectrum 
(see figure \ref{nearspec})
of the horn-off sample allows 
almost complete rejection of the $\nu_\mu$ charged current 
 events because the muons tend to be longer.  
 These data are  used in conjunction with the standard low energy 
 beam configuration data 
(horn-on) to extract the individual NC and CC-\numu{}
 components of the samples as a function of reconstructed energy. 
The Monte Carlo is used to calculate the ratios $r_{NC}$ and $r_{CC}$, 
which are 
the ratios 
of the horn-off to horn-on configurations for NC or CC events, respectively.   
An additional input from the Monte Carlo is the small contamination of 
$\nu_e$ events 
in the beam. With this information a calculation is performed for 
every energy bin to 
extract the CC and NC composition in both horn on and horn off spectra. 
Figure \ref{honoff} shows the final 
result for the ANN selection.  Integrating over the energy spectrum, 
the ND background is 57$\pm$5\% NC, 
32$\pm$7\% \numu{}-CC and 11$\pm$3\% intrinsic beam \nue{}-CC events. 
The errors on the NC and \numu{}-CC components arise from the statistics of the 
horn-off data and systematics on the ratios;
the uncertainty on the beam \nue{}-CC 
includes  systematic errors from the beam flux, 
cross-section and selection efficiency for electrons.

\begin{figure}
\centering\leavevmode
\mbox{\includegraphics[angle=0,width=0.49\textwidth]{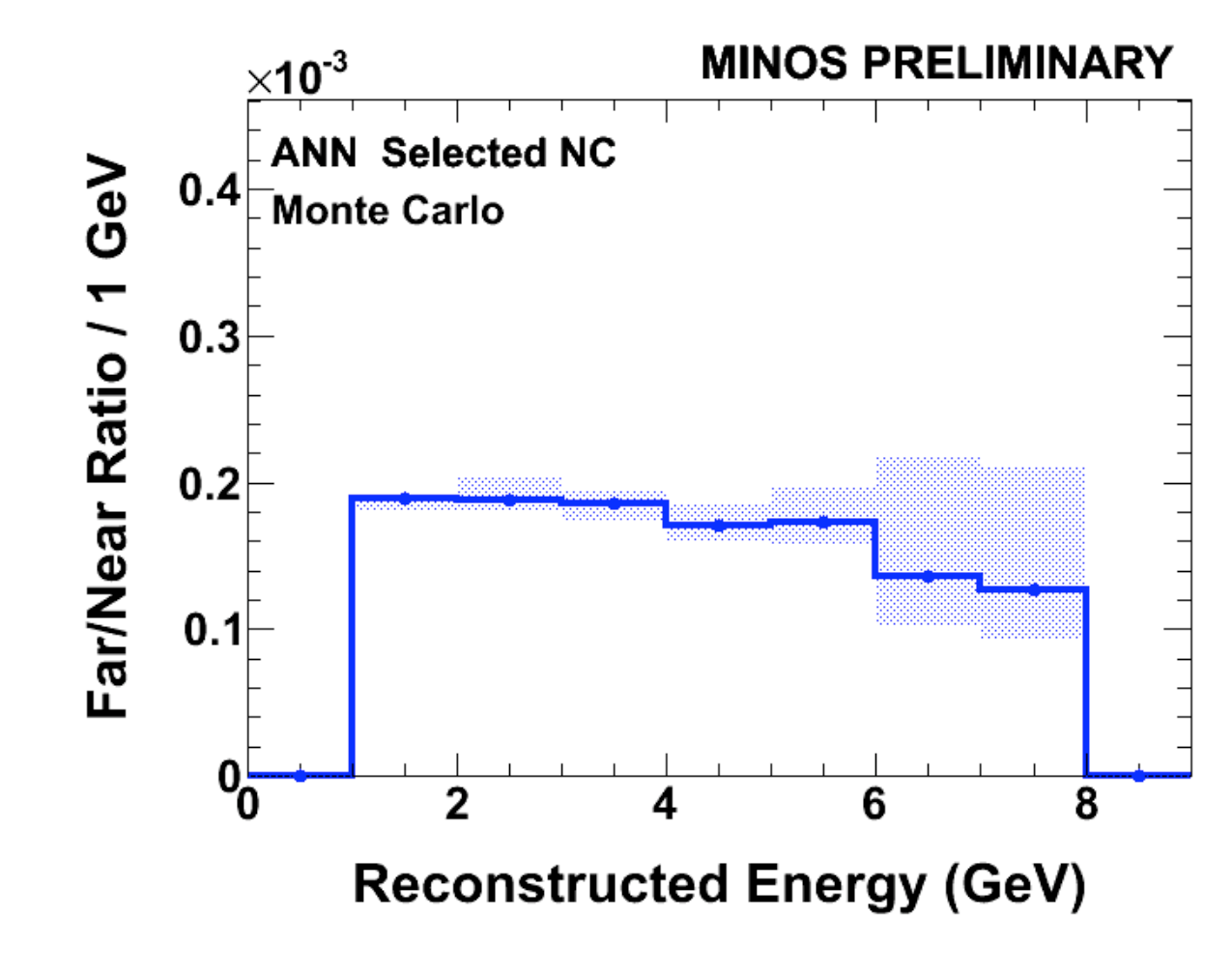}
\includegraphics[angle=0,width=0.49\textwidth]{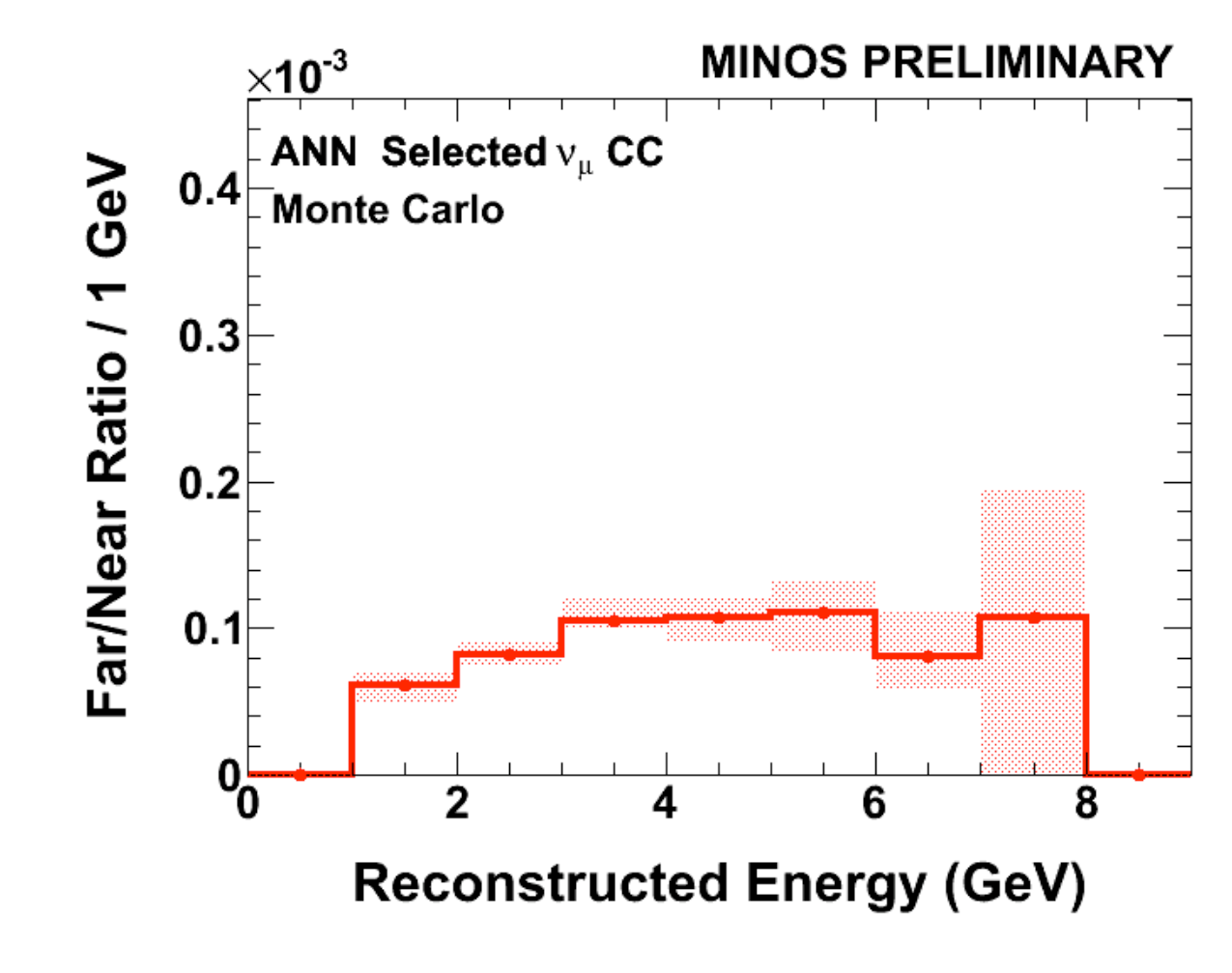}}
 \caption{(in color)
Monte Carlo calculation of the far/near ratios for NC and CC background 
components. The calculation includes effects of beamline geometry (including 
the $1/r^2$ loss), fiducial mass, difference in spetra, detector calibrations,
and differences in the analysis efficiencies.  
Plots are similar for the alternate selection (LEM).  
  \label{nf} }
\end{figure}

After decomposing the Near Detector energy spectrum into its background components, 
each background spectrum is multiplied by the ratio of the Far to Near ratio from 
the MC simulation for each component to provide a prediction of the FD spectrum for that component.
The far/near ratios are shown in figure \ref{nf} for the ANN selection.
   The MC simulations take into account differences in the spectrum of events at the
 ND and FD due to the beam line geometry as well as possible differences in detector
 calibrations and topological response. 
  Oscillations are included when predicting the 
\numu{}-CC component.  The smaller \nutau{}-CC and beam \nue{}-CC components are 
calculated by Monte Carlo using the expected energy spectrum in the FD.
  All background components are then 
added together and summed over the energy range to provide the total predicted background 
in the Far Detector. 
The detailed modeling of all  
far/near differences change the background prediction from the lowest order by  $\sim$10\%. 
 We expect a total background of 26.6 events for the ANN selection,
 of which 18.2 are NC, 5.1 are \numu-CC, 2.2  are beam \nue{} and 1.1 are \nutau{}
 for $3.14\times10^{20}$ POT \footnote{Using \dmsqx{32}=$2.43\times10^{-3} {\rm eV^{2}}$,
$ \sin^2 2 \theta_{23}=1.0$, and $\sin^2 2 \theta_{13}=0.$}. With LEM, we expect 21.4 
background events, with 14.8 NC, 2.9 \numu{}-CC, 1.1 beam \nue{} and 2.7 \nutau{}.

The effects of systematic errors were evaluated by generating modified MC samples, 
and quantifying the change in the number of predicted background events in the Far 
Detector using Far to Near ratios from the modified samples relative to the unmodified case. 
 % Figure~\ref{fig:systematics} shows the main systematic components evaluated and the 
%percentage change in the number of predicted background events expected with the ANN selection.   
Many uncertainties, including those that affect neutrino interaction physics, shower hadronization, 
intranuclear re-scattering, and absolute energy scale errors affect the events in both detectors in 
a similar manner and largely cancel.  Other effects give rise to Far/Near 
differences such as relative event rate normalization, calibration errors, reconstruction 
differences between the detectors and low level modeling of each detector. 
 The individual systematic errors are added in quadrature along with the 
systematic error arising from the decomposition of the background sources
 in the ND to give an overall systematic error of 7.3\% on the number of
 background events selected with the ANN selection. 
The LEM selection is
 more sensitive to uncertainties in the PMT gains, relative energy calibration and crosstalk.
  The total systematic error on the number of background events 
selected by the LEM technique is 12.0\%.

\subsection{Examination of events outside the signal region and other checks} 

Three main checks are performed by utilizing an independent data set obtained from 
$\nu_\mu$ charged current events in which the muon is removed in software and the 
remaining hadronic shower is analyzed as if it is a complete neutrino event.   
 This procedure is carried out on data and MC and the \nue{} selections are applied to both.  
The discrepancy between muon-removed data and MC simulation is similar to that found in the 
standard sample as a function of reconstructed energy  and  of many different 
reconstructed shower topology variables used in the selections. 

In the first check, 
the muon-removed data is used  to obtain the relative 
contributions of the background components present in the ND data spectrum. 
 The number of selected NC events in the MC simulation of the standard sample
 is scaled in each energy bin by the ratio of the number of events in the 
muon-removed data to the muon-removed simulation.  Once the number of NC events is 
determined, the number
 of \numu-CC events selected in each reconstructed energy bin in the data is
 determined using 
$N_{CC} = N^{data}_{total} - N_{NC} - N_{\nu_e}$, in which  
  the number of beam \nue{} events are
obtained from the MC. 
 The background components as derived from the muon-removed sample agree 
well with those obtained from the horn-off method.  

In the second check, we treat the muon removed data from the near and far detectors
as if they are real events and perform a complete analysis. From the near data we 
create a prediction for the far detector and count the number in the far detector. 
Using this procedure we predicted $29\pm 5(stat) \pm 2(syst)$ events for the ANN selection 
and observed 39 events. For the LEM selection the prediction was $17\pm 4(stat) \pm
2(stat)$ and the observation was 25. The observed  excess in this 
 sample, which contains no electron signal events, 
 was a cause of concern, however, upon examination of the 
full distribution with and without the particle ID cut, 
it was considered  likely to be a  statistical fluctuation. 
This issue will be explored with the larger data sample being acquired at this time.

In the third check, we estimate the efficiency for selecting \nue{}-CC events. 
 We use the sample of muon removed events and embed a simulated 
electron of the same momentum as the removed muon. 
 Test beam measurements indicate that electrons are well 
simulated in the MINOS detectors. 
Data from the test beam \cite{ref:caldet} was analyzed using the same 
selection cuts and agrees with Monte Carlo within 2.6\% for ANN and 2.2\% for LEM. sxs
 Comparisons between muon-removed data and 
simulated samples of events with embedded electrons
 indicate that the selection efficiency of \nue{} signal events is 
well modeled by the MC. The algorithms focus
 on the EM core of the shower and are 
not affected by hadronic shower modeling discrepancies.
 The difference between the data and the MC is used as a
 correction to the signal selection efficiency and
 it is -0.3\% for ANN and -5.3\% for LEM. The selection efficiency 
of the ANN selection is calculated to be 41.4$\pm$1.4\% and for
 LEM is 45.2$\pm$1.5\%.

The prediction of the backgrounds in the FD and the systematic
 uncertainties on that prediction were established before examining the data in the FD. 
Some additional checks were performed before opening the signal region.  
  The number of events passing the preselection cuts,
 but failing the \nue{}-CC selection cuts were compared to the  expectation.  
In the FD data, 146 events were observed below the ANN selection cut, 
with an expectation of $132\pm12{\rm (stat)}\pm8{\rm(syst)}$. 
 The events below the LEM cut totaled 176 events compared to
 an expectation of $157\pm13{\rm (stat)}\pm3{\rm(syst)}$. 
 Both observations deviate from the background prediction by 
approximately $1\sigma$ assuming no signal events in this part of the 
data.

\section{Results for $\nu_\mu \to \nu_e$} 

\begin{figure}
\centering\leavevmode
\mbox{\includegraphics[angle=0,width=0.49\textwidth]{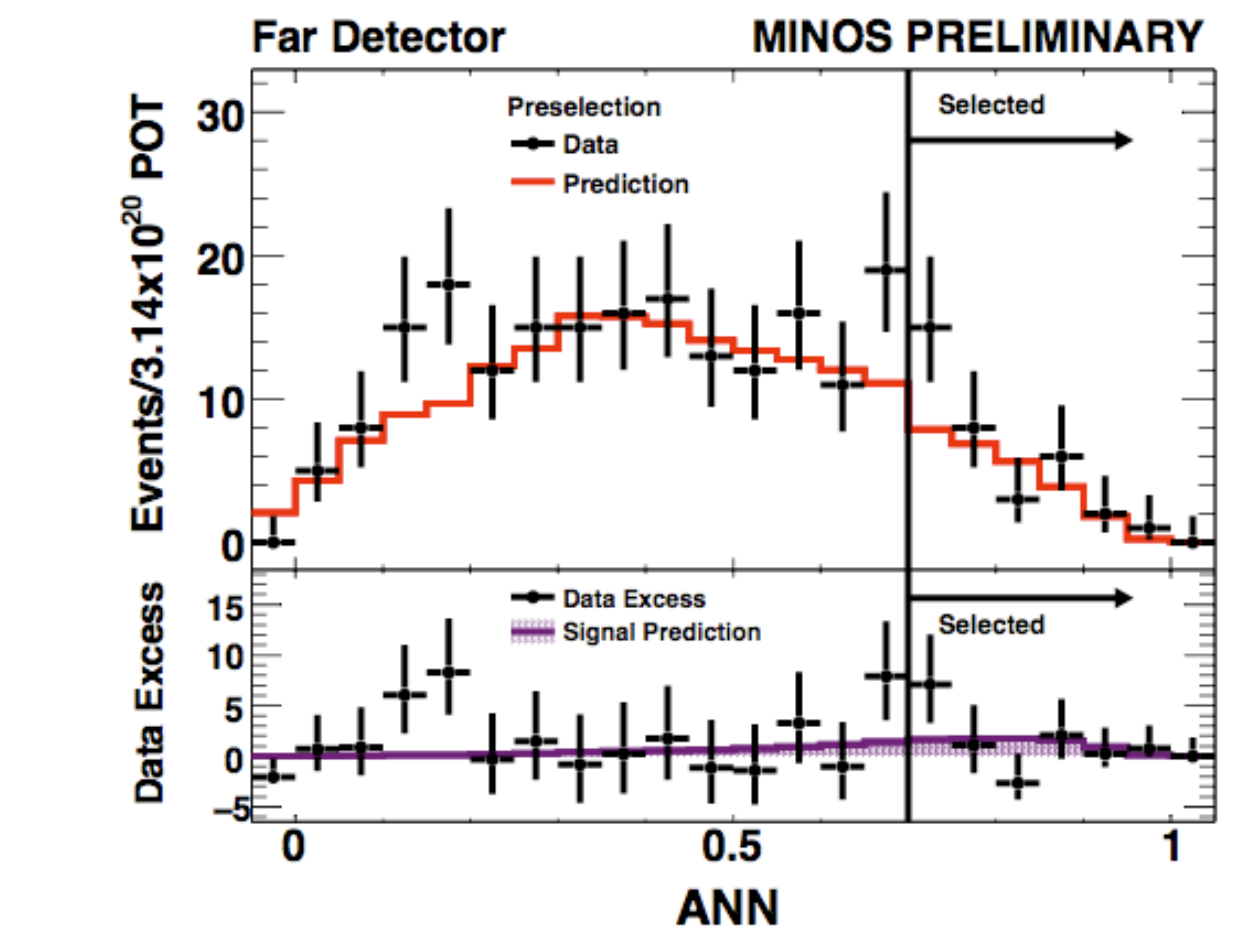}
\includegraphics[angle=0,width=0.49\textwidth]{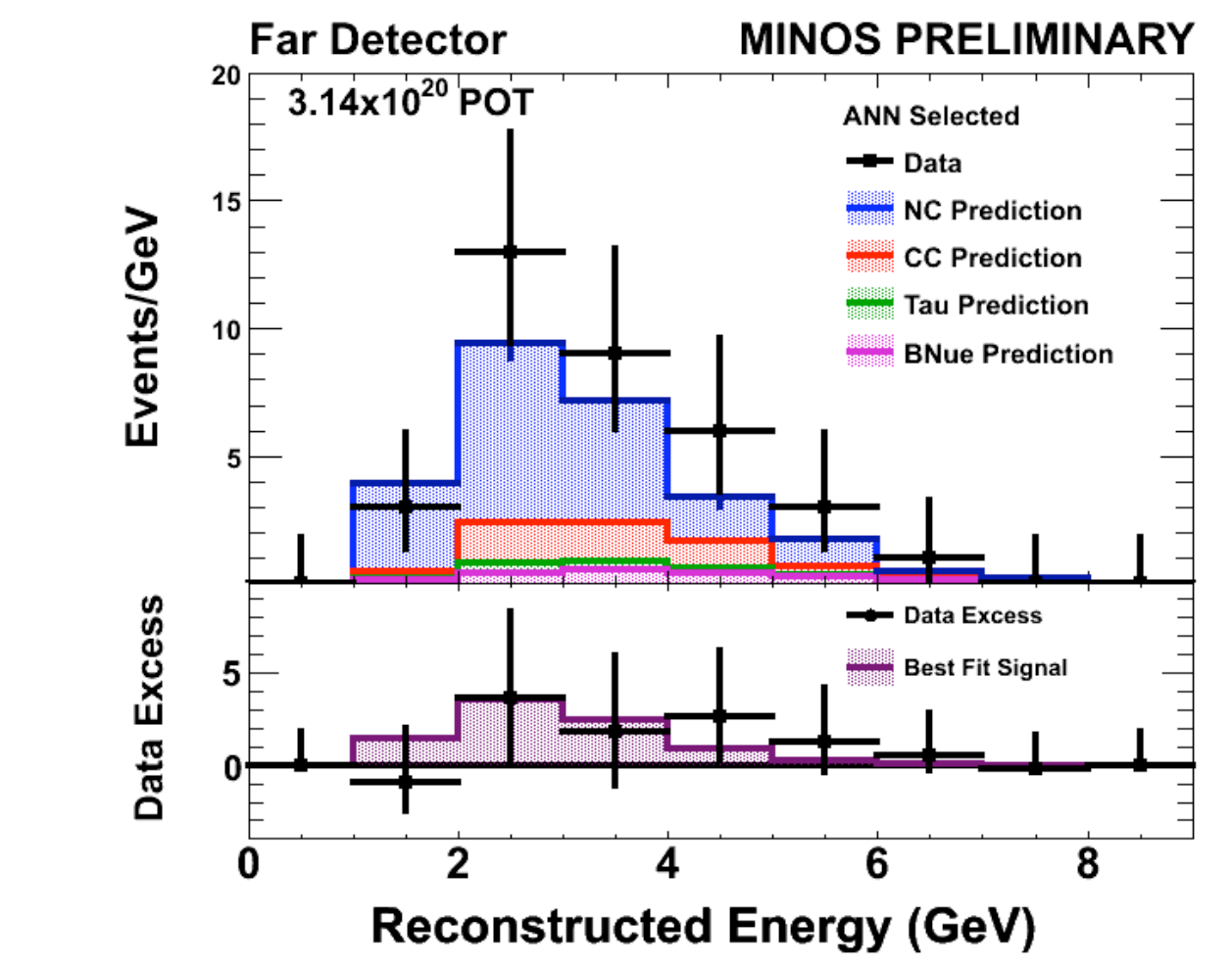}}
 \caption{(in color)
Distribution of far detector events for the ANN PID. Left shows the 
ANN PID distribution. Right shows the energy distribution after the 
PID cut. The plots below show the data minus the background prediction
with the expected distribution of the signal if all the excess is interpreted 
as signal. 
  \label{fig:fdpid1} }
\end{figure}

\begin{figure}
\centering\leavevmode
\mbox{\includegraphics[angle=0,width=0.49\textwidth]{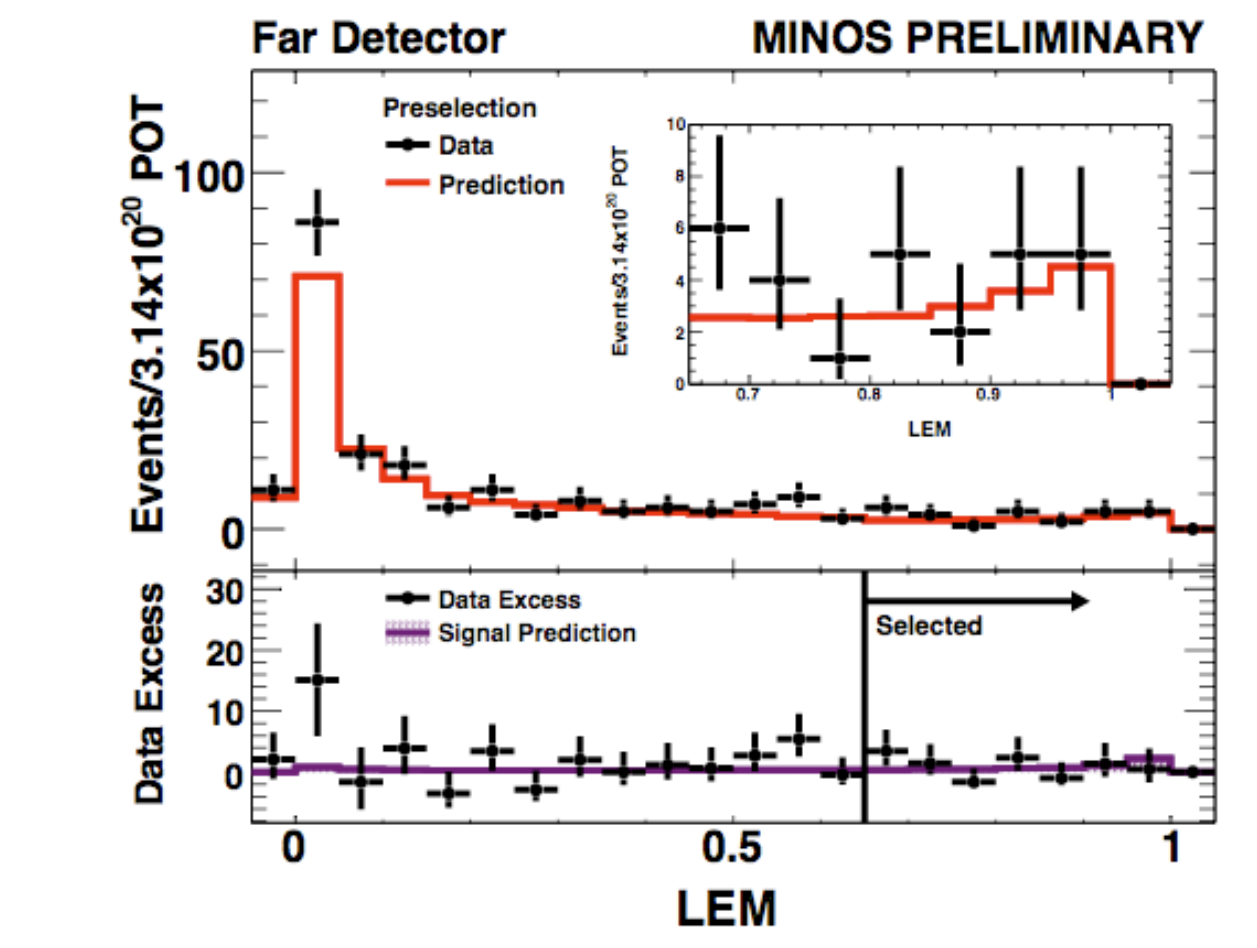}
\includegraphics[angle=0,width=0.49\textwidth]{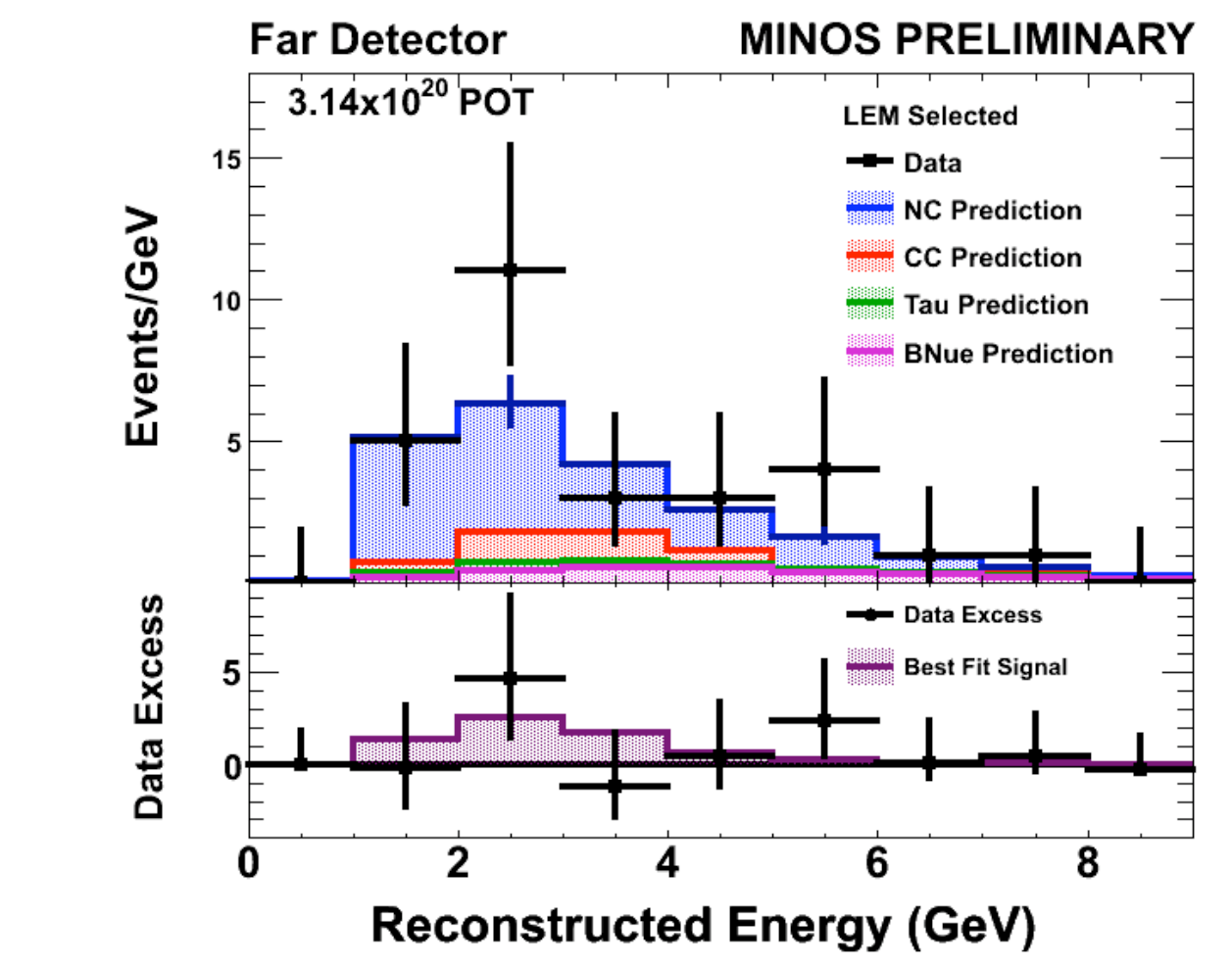}}
 \caption{(in color)
Distribution of far detector events for the LEM PID. Left shows the 
LEM PID distribution. Right shows the energy distribution after the 
PID cut. The plots below show  the data minus the  background prediction
with the expected distribution of the signal if all the excess is interpreted 
as signal. 
  \label{fig:fdpid2} }
\end{figure}

After examining the sideband data sets, we proceeded to count the
 number of events passing the predetermined selection cut.  
We observe  \annobs{} events in the FD when using the \nue{} 
selection based on the ANN algorithm, with a background expectation of 
 \annpred{}.  With LEM (the secondary selection) we
 observe \mcnnobs{}, with a background expectation of \mcnnpred{}. 
 The  distributions are shown in Fig.~\ref{fig:fdpid1} and \ref{fig:fdpid2}.

\begin{figure}
\begin{center}
\includegraphics[angle=0,width= 0.7 \textwidth]{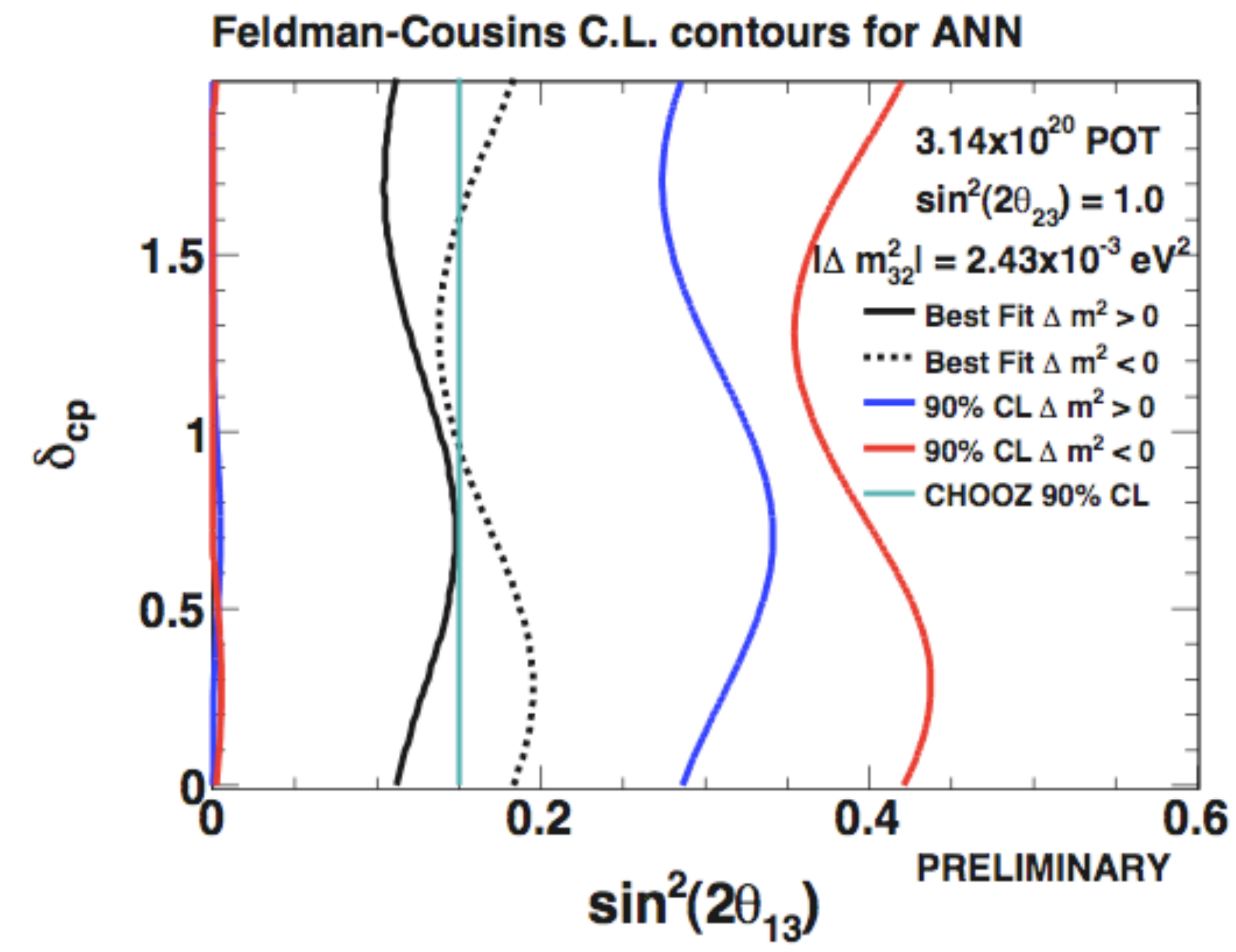}
\end{center}
\caption{
The 90\% confidence interval in 
the  \sinsq{13} and $\delta_{CP}$ plane
using the ANN PID results.   Black lines show the best fit
 to our data in both the normal hierarchy (solid) and inverted hierarchy (dotted). 
 Blue (red) lines show the 90 C.L. boundaries for the normal (inverted) hierarchy. }
\label{fig:sens}
\end{figure}

 Figure~\ref{fig:sens} shows the 90\% confidence level interval 
in the  $\sin^2 2 \theta_{13}$ and $\delta_{CP}$ plane for each mass hierarchy
using our observation for ANN PID.
To set this limit we have used only the total observed number of events; 
detailed 
fitting of the data distributions was not performed for this result. 
 We use the current best fit value of
 $|\delmsq{32}|$=\unit[2.43$\times 10^{-3}$] {${\rm eV^{2}}$} and  \sinsq{23}=1.0 for 
this calculation.   
Fluctuations (Poisson) and systematic
 effects (Gaussian) are incorporated via
 the  Feldman-Cousins approach \cite{ref:pdg}. 
The oscillation probability is computed using a 
full 3-flavor neutrino mixing framework that includes
 matter effects.

\section{Updates to MINOS measurement of $\nu_\mu$ disappearance} 

\begin{figure}
\centering\leavevmode
\mbox{\includegraphics[angle=0,height=0.49\textwidth]{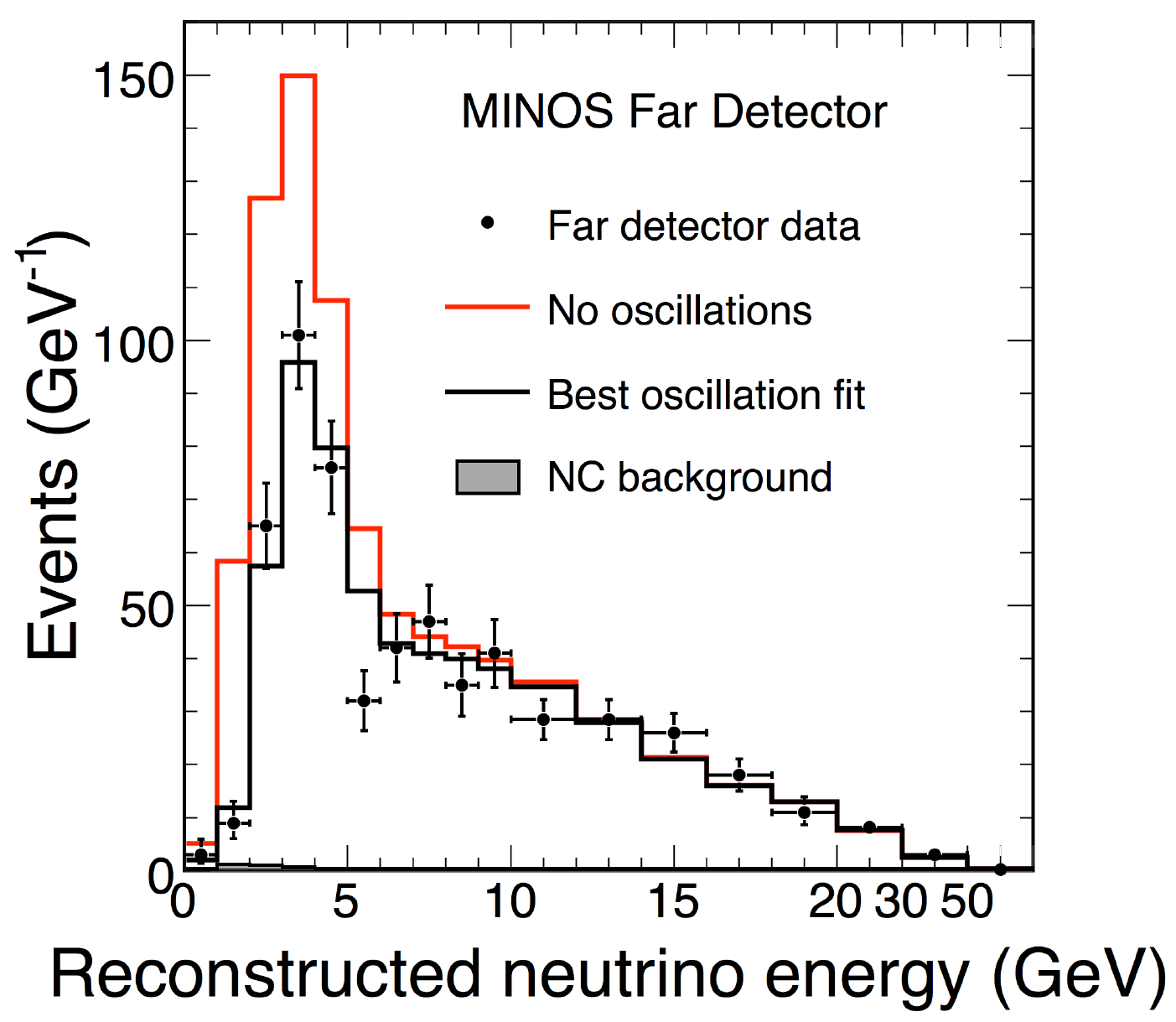}
\includegraphics[angle=0,height=0.49\textwidth]{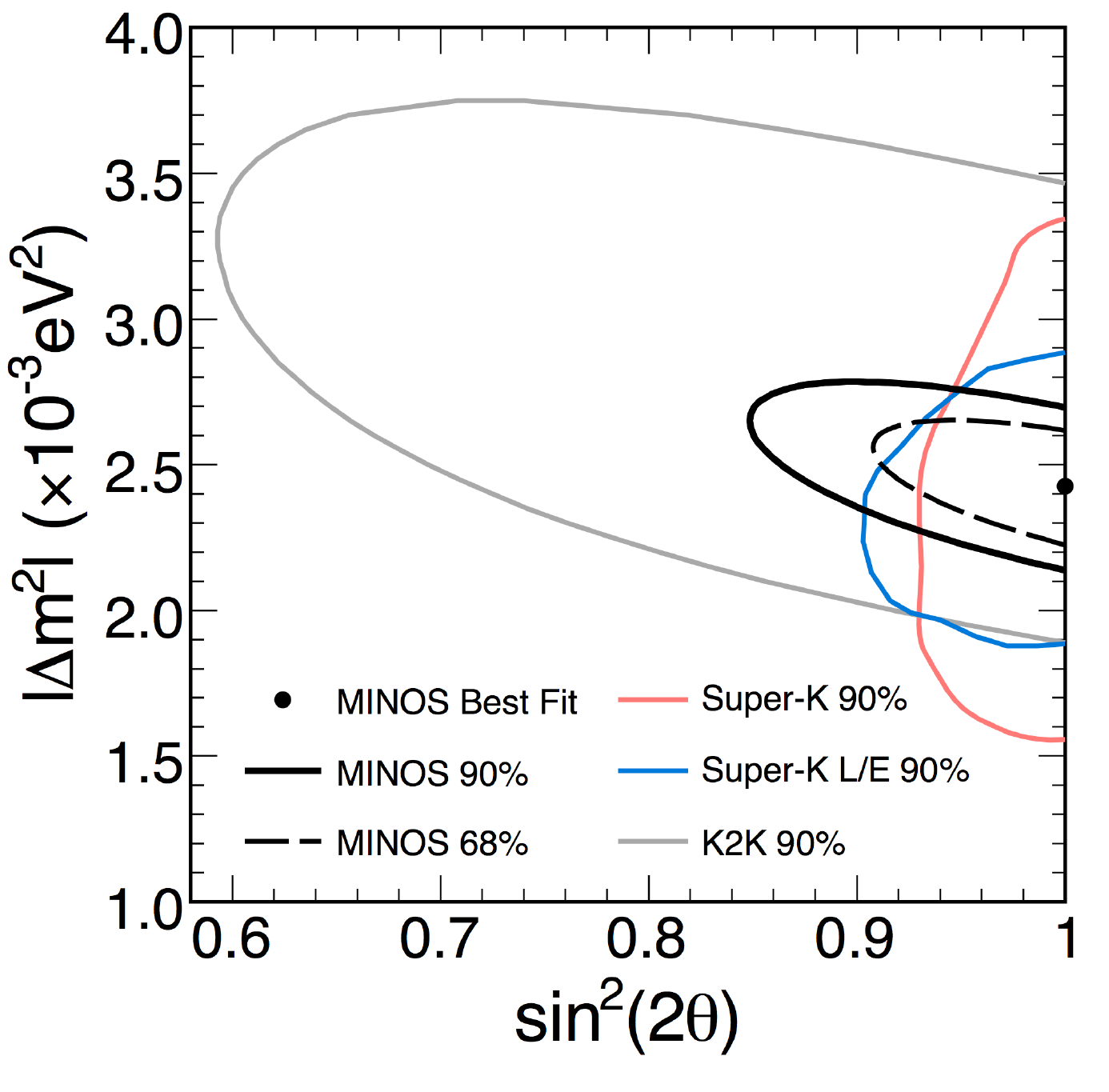}}
 \caption{(in color)
Measurement of MINOS for the disappearance of muon neutrinos. 
  \label{disap} }
\end{figure}

MINOS has recently reported updated measurements of $\nu_\mu$ 
disappearance \cite{ref:minosprl2} on the same data set that 
was described above (the high energy spectrum data was also 
included).  We observed 848 $\nu_\mu$-CC events in the 
far detector across the energy range of 0 to 120 GeV compared to 
the expectation of $1065\pm 60 (syst)$. The observed spectrum is 
shown in figure \ref{disap}. The same figure also shows the 
confidence interval in the $\sin^2 2 \theta_{23}$ versus 
\dmsqtwo ~plane.  We obtain $|\Delta m^2 _{32}| = 2.43 \pm 0.13 \times 
10^{-3} eV^2$ at 68 \% C.L.  and the mixing angle  of 
$\sin^2 2 \theta_{23} > 0.90$ at 90\% C.L.  At present time 
the  measurement of \dmsqtwo ~is dominated by MINOS 
and the  measurement of the mixing angle is dominated by 
Super-Kamiokande. 

\begin{figure}
\centering\leavevmode
\includegraphics[angle=0,width=0.6\textwidth]{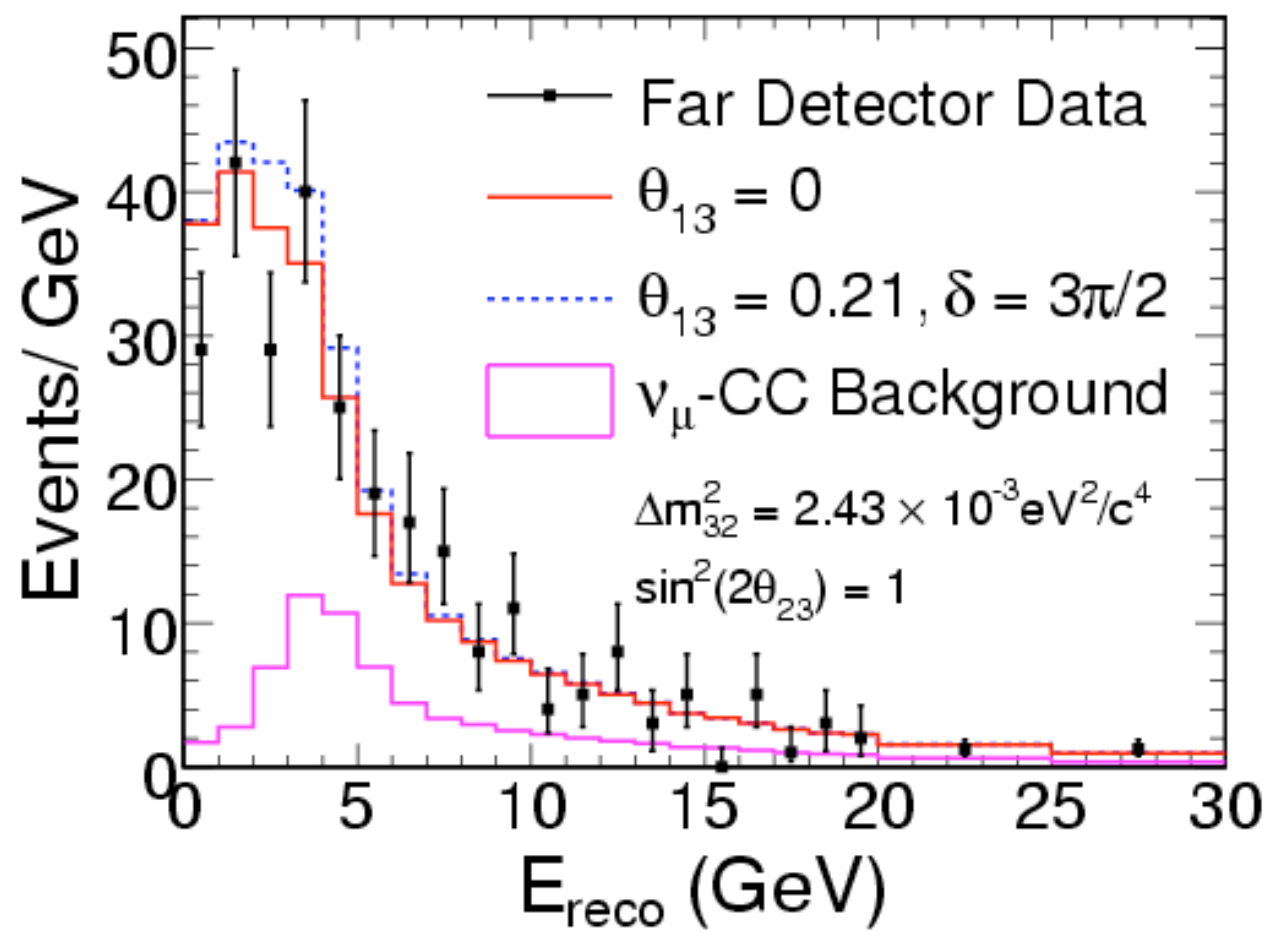}
 \caption{(in color)
Measurement of Neutral Current spectrum in MINOS.
  \label{nc} }
\end{figure}

While the disappearance of $\nu_\mu$ from the atmosphere and the NuMI/MINOS
beam experiment is largely explained by 3 generation neutrino mixing with
$\nu_\mu \to \nu_\tau $ as the mechanism, any small admixture of sterile neutrinos 
is still an experimental issue.  MINOS performed a search for disappearance of active
neutrinos using neutral current interactions \cite{ref:minosnc}.  
The final spectrum of neutral current events and the prediction based on 
near detector data is shown in figure \ref{nc}. 
No anomalous depletion in the reconstructed energy 
spectrum is observed. Assuming oscillations occur
 at a single mass-squared splitting, a fit to the
neutral- and charged-current energy spectra
 limits the fraction of $\nu_\mu$
 oscillating to a sterile neutrino to be below 0.68 at 90\% confidence level.
Electron neutrinos can constitute a background to the neutral current analysis, 
therefore any possible contribution from $\nu_e$ appearance at the current experimental
bound leads to a less stringent limit.

\section{Conclusions} 

In summary, we report the first results of a search for 
\nue{} appearance in the MINOS experiment. 
The observed rate of events in the Far Detector after \nue{} selection for 
\pot{} is consistent with the background expectation within 1.5 standard 
deviations. 
For this data set, assuming 
 $|\delmsq{32}|$=\unit[2.43$\times 10^{-3}$] {${\rm eV^{2}}$}, 
  \sinsq{23}=1.0, and  $\delta_{CP}=0$, we set an upper limit of  
$\sin^2(2\theta_{13})< 0.29$ at 90\% C.L. for the normal hierarchy 
and $\sin^2(2\theta_{13})< 0.42$ for the inverted hierarchy.

\section{Acknowledgements}
This was prepared for the proceedings of the 
XIII International Workshop on Neutrino Telescopes at the Istituto Veneto 
di Scienze, Lettere ed Arti in Venice held on March 10-13, 2009.  
The presentation was on behalf of the MINOS collaboration. 
This work was supported by the US Department of Energy under contract 
number DE-AC02-98CH10886.

\end{document}